\documentclass[useAMS,usenatbib,onecolumn]{mn2e}
\usepackage{epsfig,amssymb}
\newcommand{\be}{\begin{equation}}
\newcommand{\ee}{\end{equation}}
\usepackage{graphicx}

\title[Evolution of force-free magnetic fields in a magnetar]{Nonideal evolution of nonaxisymmetric, force-free magnetic fields in a magnetar}
\author[A. Mastrano and A. Melatos]{A. Mastrano$^{1}$\thanks{E-mail:
amastran@physics.unimelb.edu.au} and A.
Melatos$^{1}$\thanks{E-mail: amelatos@physics.unimelb.edu.au}\\
$^{1}$School of Physics, University of Melbourne, Parkville VIC
3010, Australia}

\begin{document}

\date{Accepted ?. Received ?; in original form ?}

\pagerange{\pageref{firstpage}--\pageref{lastpage}} \pubyear{?}

\maketitle

\label{firstpage}

\begin{abstract}

\noindent{Recent numerical magnetohydrodynamic calculations by
Braithwaite and collaborators support the `fossil field' hypothesis
regarding the origin of magnetic fields in compact stars and suggest
that the resistive evolution of the fossil field can explain the
reorganisation and decay of magnetar magnetic fields. Here, these
findings are modelled analytically by allowing the stellar magnetic
field to relax through a quasistatic sequence of nonaxisymmetric,
force-free states, by analogy with spheromak relaxation experiments,
starting from a random field. Under the hypothesis that the
force-free modes approach energy equipartition in the absence of
resistivity, the output of the numerical calculations is
semiquantitatively recovered: the field settles down to a linked
poloidal-toroidal configuration, which inflates and becomes more
toroidal as time passes. A qualitatively similar (but not identical)
end state is reached if the magnetic field evolves by exchanging
helicity between small and large scales according to an
$\alpha$-dynamo-like, mean-field mechanism, arising from the
fluctuating electromotive force produced by the initial random
field. The impossibility of matching a force-free internal field to
a potential exterior field is discussed in the magnetar context.}

\end{abstract}


\section{Introduction}

Magnetars are neutron stars whose (un)pulsed X-ray luminosities
($\lesssim 10^{29}$ W) exceed their spin-down luminosities
\citep{TD95}. They fall into two classes: soft gamma repeaters
(SGRs) \citep{Retal94,Hetal05} and anomalous X-ray pulsars (AXPs)
\citep{TD96,M00,MLetal03,WT06}. From their observed spin periods and
spin-down rates, one can infer that magnetars possess very high
surface magnetic fields $B \geqslant 5 \times 10^9$ T\footnote{Throughout this paper, we use SI units. 1 T $= 10^{4}$ G.}, higher than
most (although not all) rotation-powered pulsars
\citep{Ketal98,M99,KML05}. Duncan and Thompson (1992) proposed that
the resistive decay of the magnetic field powers the continuous
X-ray emission from AXPs and the intense outbursts (lasting a
fraction of a second) from SGRs, e.g. Palmer et al. (2005). AXPs may
be SGRs which are yet to burst \citep{TD96,M00}.

It remains unclear how such strong stellar magnetic fields come to
exist. Broadly speaking, there are two hypotheses for their origin
\citep{BS06}: (1) fossil field, where the progenitor star's magnetic
field is amplified by compression during core collapse and is frozen
into the highly conducting compact remnant; and (2) dynamo
amplification, where the field is generated by a convective dynamo
in the proto-neutron star. (In this paper, we discuss dynamos operating in the protostar stage rather than continuously regenerating the magnetic field of a neutron star, white dwarf, or Ap star during the star's lifetime; the latter idea seems to be ruled out by observations of Ap stars and white dwarfs, as we discuss below.) Observations to date do not favour conclusively one hypothesis over the other. Indeed, the hypotheses are not mutually exclusive; a dynamo operating in the protoneutron star/white dwarf stage may amplify the fossil field inherited from the main sequence. Fossil fields are
supported by the observed magnetic field distribution in Ap stars
and white dwarfs \citep{FW06}. On the other hand, the magnetar birth
rate implied by observations agrees with the operation of a
large-scale dynamo in rapidly rotating, massive proto-neutron stars
(with masses $\gtrsim 25 M_{\sun}$ and periods $\sim 1$ ms)
\citep{Ketal94,Getal99,Hetal03,Getal05,PP06}. On the theoretical front, one of the major issues with the fossil field hypothesis is the (in)stability of the progenitor field. Much work has been done to show that purely poloidal and purely toroidal field configurations are unstable (Prendergast 1956; Wright 1973; Goossens \& Tayler 1980; van Assche, Tayler, \& Goossens 1982), but a linked toroidal-poloidal configuration may be stable (Tayler 1980).

There are, in fact, two different fossil field hypotheses: ``strong", where the magnetic field of the progenitor is $\sim 1 $ T for a $1 M_\odot$ star, following compression of a strongly magnetised ($3\times 10^{-2}$ T) molecular cloud (Moss 2001, 2003), and ``weak", where the field is generated after the progenitor forms by various dynamo processes [see e.g., Charbonneau \& MacGregor (1996) and Spruit (2002)]. In this paper, we do not distinguish between the two hypotheses, as our analysis is too elementary to treat the magnetization of the progenitor. Magnetic flux conservation implies that a magnetar progenitor needs to have magnetic field strength $B\gtrsim 1$ T (Woltjer 1964; Vink \& Kuiper 2006). The core and envelope of the progenitor are strongly coupled, angular momentum is rapidly transferred away by a stellar wind, and the core rotates rather slowly, for example $P\sim 10$ ms for 15 $M_\odot$ stars and $P\sim 4$ ms for 35 $M_\odot$ stars (Spruit 2002; Heger, Woosley, \& Spruit 2005). This is slower than the $P\sim 1$--$3$ ms needed for an efficient dynamo to operate (Duncan \& Thompson 1992; Thompson \& Duncan 1995, 1996; Heger et al. 2005; Vink 2008). Heger et al. (2005) calculated the effects of magnetic braking on newly born neutron stars and concluded that $P\lesssim 3$ ms is achieved only in stars with $M\gtrsim 25 M_\odot$ (Gaensler et al. 2005; Heger et al. 2005). The requirement of rapid rotation suggests a stringent absolute test of the dynamo hypothesis: are enough neutron stars born with such rapid rotations to explain the observed magnetar birth rate? Taking $25 M_\odot$ to be the mass cutoff, Gaensler \emph{et al.} (2005) estimated the birth rate of magnetars to be $\sim 10 \%$ that of radio pulsars, consistent with current populations of AXPs and SGRs (Kouvelioutou et al. 1994; Gaensler et al. 1999). In contrast, the explosion energies of the supernova remnants associated with the AXPs 1E 1841--045 (Kes 73) and 1E2259+586 (CTB109) and the SGR 0526--66 (estimated from the shock speed, which is determined from X-ray spectra measured by \emph{XMM-Newton}) suggest an initial spin period $P\gtrsim 5$ ms, not enough for a dynamo (Vink \& Kuiper 2006; Vink 2008).

    The lack of correlation between the observed field strength and spin periods of Ap stars and white dwarfs seems to argue against the hypothesis that a dynamo can operate in the core to regenerate the magnetic field continuously (Braithwaite \& Spruit 2004; Petit et al. 2007). Observations of main sequence stars that are potential magnetar progenitors do not conclusively favour either fossil fields or protostar dynamos. From the similar fluxes of magnetic A stars (called Ap stars) and white dwarfs, Ferrario and Wickramasinghe (2006) inferred that O and B progenitors need to have dipolar fields of strength $10^{-4}$ T $\lesssim B \lesssim 10^{-1}$ T to generate neutron stars with $10^7$ T $\lesssim B\lesssim 10^{11}$ T. Unfortunately, the magnetic fields of these O and B stars are hard to observe. This is because their absorption lines in the optical range are few and broad (perhaps due to rotation), making the Zeeman effect, by which the magnetic fields are measured, harder to detect (Donati et al. 2006a, b; Vink 2008). Two magnetic stars, HD 191612 (O class) and $\tau$ Sco (B class), are observed to have fairly long periods, 41 and 538 days respectively, which are inconsistent with the dynamo hypothesis (Donati et al. 2006a, b). Note, however, that their current rotation periods may not reflect directly on their periods during the protostar phase [many processes, such as accretion disk locking, can slow down the star after the stable field is in place (Braithwaite 2008)]. So, while the magnetic field today may not be generated continuously by a dynamo, it may arise from a dynamo  during the protostar stage. It must also be kept in mind that our knowledge of magnetism in main sequence stars of masses $> 8M_\odot$ is sketchy.

Recent magnetohydrodynamic (MHD) simulations have shown that an unstable, random initial field can decay into a long-term stable configuration \citep{BN06,BS04,BS06}. Starting from a
random magnetic field, the simulations follow the field as it
relaxes towards a lower energy state, with and without diffusion,
transforming spontaneously into a stable configuration: a nearly
axisymmetric torus with roughly equal poloidal and toroidal fluxes,
which expands as a function of time. Poloidal field lines are
wrapped around the torus, extending through the atmosphere and
creating an `approximate dipole' \citep{BN06}. It must be noted that the results of the cited numerical simulations do not say anything directly about whether a neutron star's field comes from a dynamo process operating during the protoneutron star phase or simply conserved from the main sequence progenitor. A stable magnetic field configuration is necessary in both fossil field and dynamo hypotheses.

Motivated by these numerical simulations, we ask here whether it is
possible to model the magnetic evolution approximately yet
analytically as a slow, helicity-conserving relaxation through a
quasistatic sequence of force-free states, by analogy with spheromak
experiments. A previous paper on relaxing force-free magnetic fields
focused on axisymmetric modes \citep{BN07}. In this work, we draw
particular attention to the importance of nonaxisymmetric modes,
without which it is impossible to reproduce the linked
poloidal-toroidal field lines observed in the simulations
\citep{BN06,BS04,BS06}. We also allow our model neutron star to
develop a small-scale turbulent velocity field as the magnetic field
relaxes, generalizing Broderick and Narayan's (2007) work to
accommodate (some of) the physics of a mean-field dynamo
\citep{BF04}.

In Section 2, we delineate our aim and the four hypotheses regarding
the nondiffusive evolution of the force-free field that we
investigate. In Section 3, we discuss the basic properties of
force-free magnetic fields in detail. Then, in Section 4, we discuss
the results of our study of the nondiffusive evolution of a
composite force-free field through the four different hypotheses. In
Section 5, we briefly discuss the diffusive evolution of a
force-free field. Finally, in Section 6, we present our conclusions
and discuss in greater detail the fundamental problem of matching a
force-free magnetic field to a potential field.

\section[]{Evolutionary Hypotheses}

The aim of this paper is to construct a simple, \emph{approximate}, analytic model for the evolution of magnetar fields (via non-diffusive and diffusive processes), which reproduces coarsely the numerical results of Braithwaite and collaborators. In pursuing this objective, we hypothesize that the
field evolves according to two principles: (1) it passes
quasistatically through a series of force-free equilibria, each a
superposition of axisymmetric and nonaxisymmetric modes; and (2) the
relative weightings of the modes are controlled by the conservation
(or slow evolution) of global quantities like magnetic helicity,
energy, or hydrodynamic vorticity. Obviously, this approach
sacrifices realism for simplicity; it is emphatically not a
substitute for high-resolution, resistive MHD simulations.

We choose force-free fields as our basis because of their
astrophysical prevalence \citep{P84} and analytic simplicity. This
choice is justified by laboratory experiments with spheromaks, which
are topologically congruent with a star, where it is observed that a
spherical magnetic field relaxes into a force-free axisymmetric
state, known as the Taylor state
\citep{RB79,Getal80,Yetal81,Jetal85,T86}. In a magnetar, the
force-free approximation is defensible with respect to latitudinal
and azimuthal force components, with $\nabla p \ll
\mu_0^{-1}(\nabla\times {\bf{B}})\times {\bf{B}}$ in these
directions (where $p$ is the pressure and ${\bf{B}}$ is the magnetic
field). The approximation breaks down in the radial direction in a
realistic star, due to the strong gravitational force. However, the
main aim of this paper is to make contact with the MHD simulations
of Braithwaite and collaborators. For this
purpose, as noted below, the force-free approximation is adequate
overall.

We stress here that nothing physical requires the field to be force-free; indeed, with $\nabla P \gg {\bf{J}}\times{\bf{B}}$, one would na\"{i}vely expect ${\bf{B}}$ to be anything but force-free (at least with regard to the radial gradients). We want to make it clear that we do not propose force-free fields as a realistic model, but rather as a convenient mathematical artifice for studying the long-term evolution of magnetic helicity and field geometries in a neutron star. That is, we take the simplifying assumption that the field is force-free, to see whether the behaviour of such a field mimics that of a more general field investigated by Braithwaite et al. Additionally:

    \begin{enumerate}

    \item Force-free fields are an analytically convenient orthonormal basis in which any interior field can be expanded. It will be worth testing empirically (from the simulation output) what the coefficients in a force-free expansion are, in view of the qualitative visual agreement with the simulation results reported in Sections 4 and 5.

    \item In many other high-$\beta$ ($|\nabla p| \gg |{\bf{J}}\times{\bf{B}}|$) plasmas, e.g., in spheromaks (Jarboe et al. 1980; Turner et al. 1983; Janos et al. 1985; Barnes et al. 1986; Knox et al. 1986), the field does relax to a force-free (minimum energy) state (Taylor 1986), even though it does not have to and one na\"{i}vely would not expect it to. It does this by tending ``thermodynamically" to a state with $\nabla p\approx 0$ and ${\bf{J}}\times{\bf{B}}\approx 0$ separately (because a magnetic field does no work in the first law of thermodynamics). The analogue here is a state with $-\nabla p+\rho\nabla\Phi\approx 0$ and ${\bf{J}}\times{\bf{B}}\approx 0$.
    \end{enumerate}

The MHD simulations cited above start from a random field. The
random field is not force free, but this is immaterial: the role
played by the initial state, in our model, is to define the initial
values of conserved (or nearly conserved) global quantities like
magnetic helicity. We focus on the three lowest poloidal orders of
the magnetic field, including nonaxisymmetric modes. Our initial
state is a sum of these modes, equally weighted to emulate a random
field (i.e., spatially uncorrelated) where all modes (up to some wavenumber cutoff) are equally
probable.

Next, we entertain four possibilities for how the field evolves
\emph{without diffusion}, as in the first stage of the numerical
simulations of Braithwaite and Nordlund (2006). One possible end
state of nondiffusive evolution is energy equipartition, where each
force-free mode contributes equal energy to the total, Sec. 4.2 explains the equipartition assumption in more detail. We conjecture
that a random field may evolve into an equipartitioned state either
by keeping the total energy and helicity constant (state A) or by
keeping the total helicity constant while allowing the energy to
vary (state B). A second possible end state of nondiffusive
evolution is dictated by the dynamo theory of Blackman and Field
(2004), in which helicity is exchanged between short- and
long-wavelength modes through the agency of a turbulent
electromotive force ($\alpha$ effect). Again, the end state is
reached either by keeping the total energy and helicity constant
(state C) or by keeping the total helicity constant while allowing
the energy to vary (state D). The discretisation scheme of the simulation gives rise to a small ``effective diffusivity'', which leads to some dissipation and energy loss (by a factor of $\sim 10$ -- 100) during the initial relaxation, before a stable magnetic field configuration forms \citep{B08}. In our analysis, however, we prefer to keep an open mind and allow cases without (A and C) and with (B and D) energy loss.

Lastly, we briefly discuss how the relaxed, force-free field
diffuses over long times when the electrical resistivity is not
zero. In the numerical simulations, the magnetic field is observed
to diffuse out of the star in a roughly dipolar configuration
\citep{BN06}. We show that a field which is a linear combination of
several force-free modes diffuses while keeping the force-free
parameter $a$ (defined in Section 3) and relative weighting of the
modes constant, thus retaining its pre-diffusion shape. We also
discuss the difficulty of matching a force-free field with spatially
uniform $a$ to a potential field outside the star.

\section[]{Nonaxisymmetric Force-free fields}

Force-free magnetic fields obey the condition
$\nabla\times{\bf{B}}=a{\bf{B}}$, where $a({\bf{x}})$ (units:
inverse length) is a function of position in general. The conduction
current density, ${\bf{J}}=\mu_0^{-1}\nabla\times {\bf{B}}$, is
therefore parallel to $\bf{B}$ and the Lorentz force, proportional
to ${\bf{J}}\times{\bf{B}}$, vanishes. Hence, force-free fields
offer an attractive, static, \emph{long-term} magnetic
configuration. In general, the solution to the force-free condition
that also fulfills $\nabla \cdot {\bf{B}}=0$ can be written as
\citep{CK57,P84}

\be {\bf{B}} = a\nabla \times (\psi{\bf{n}}) +
\nabla\times[\nabla\times(\psi{\bf{n}})]\ee where ${\bf{n}}$ is some vector and $\psi$ is a scalar solution of the Helmholtz
equation

\be(\nabla^2 + a^2) \psi=0.\ee

In spherical polar coordinates $(r, \theta, \phi)$, choosing
${\bf{n}}={\bf{\hat{e}}}_r$ and restricting attention (for
simplicity) to uniform $a$, we can write the magnetic field as
\citep{P84}

\be B_r=\sum_{n=1}^{\infty}\sum_{m=0}^{n} C_{nm}
n(n+1)r^{-3/2}J_{n+1/2}(ar)P_n^m(\cos\theta)\exp{(im\phi)},\ee
\[B_\theta=\sum_{n=1}^{\infty}\sum_{m=0}^{n}
C_{nm}[ar^{-1/2}J_{n-1/2}(ar)-nr^{-3/2}J_{n+1/2}(ar)]\frac{dP_n^m(\cos\theta)}{d\theta}\exp{(im\phi)}\]
\be\phantom{++} +C_{nm}
imar^{-1/2}J_{n+1/2}(ar)P_n^m(\cos\theta)\exp{(im\phi)}/\sin\theta,\ee
\[B_\phi=\sum_{n=1}^{\infty}\sum_{m=0}^{n}
C_{nm}im[nr^{-3/2}J_{n+1/2}(ar)-ar^{-1/2}J_{n-1/2}(ar)]P_n^m(\cos\theta)\exp{(im\phi)}/\sin\theta\]
\be\phantom{++}
-C_{nm}ar^{-1/2}J_{n+1/2}(ar)\frac{dP_n^m(\cos\theta)}{d\theta}\exp{(im\phi)},\ee
where $C_{nm}$ are complex constants, $J_{n\pm 1/2}$ denotes Bessel
functions of order $n\pm 1/2$, and $P_n^m$ denotes associated
Legendre functions of the first kind. Physical field components are
the real parts of (3)--(5). We plot some of these modes in Fig. 1.
The leftmost plot in the top row shows the axisymmetric Taylor
state, the end state achieved by relaxing spheromak magnetic fields
(Taylor 1986). Note that a nonaxisymmetric vector field may have
chaotic (albeit closed) field lines (e.g. second row, centre plot,
Fig. 1) even though the vector field is smooth, because the field
line equations form a nonlinear dynamical system of dimension three
\citep{BE04}. The field lines are closed (because $\nabla\cdot {\bf{B}}=0$) but chaotic (two field lines that are arbitrarily close at some point will diverge macroscopically along their length). By contrast, the field lines of an axisymmetric vector
field must form simple loops, by the Poincar\'{e}-Bendixson theorem
for two-dimensional dynamical systems \citep{JS99}.

A nonuniform $a$ gives an extra term in the diffusion equation, since $\eta\nabla^2{\bf{B}}=-a^2\eta{\bf{B}}-\eta(\nabla a)\times{\bf{B}}$. Under the action of the new term, the field changes shape as it diffuses. In addition, a nonuniform $a$ facilitates global current closure, since the currents carried by two flux tubes of equal area (proportional to the local value of $a$) are not necessarily equal if $a$ varies from place to place. This allows return currents to be confined to thin sheets that leave most of the field undisturbed, helping greatly with the matching problem discussed in Sec. 6.

%*plots for variance 1, 1.25, 1.5, 2.0, 2.5 at z=0.362 from Monte Carlo Simulations*%
\begin{center}
\begin{figure*}
\begin{tabular}{cccc}
\begin{tabular}{c}
% (a) \\
\mbox{(1,0)}
\end{tabular}

&
\begin{tabular}{c}
% (a) \\
\mbox{(1,1)}
\end{tabular}
\\
\begin{tabular}{c}
% (a) \\
%\includegraphics[height=30mm]{1atest.eps}
\includegraphics[height=30mm]{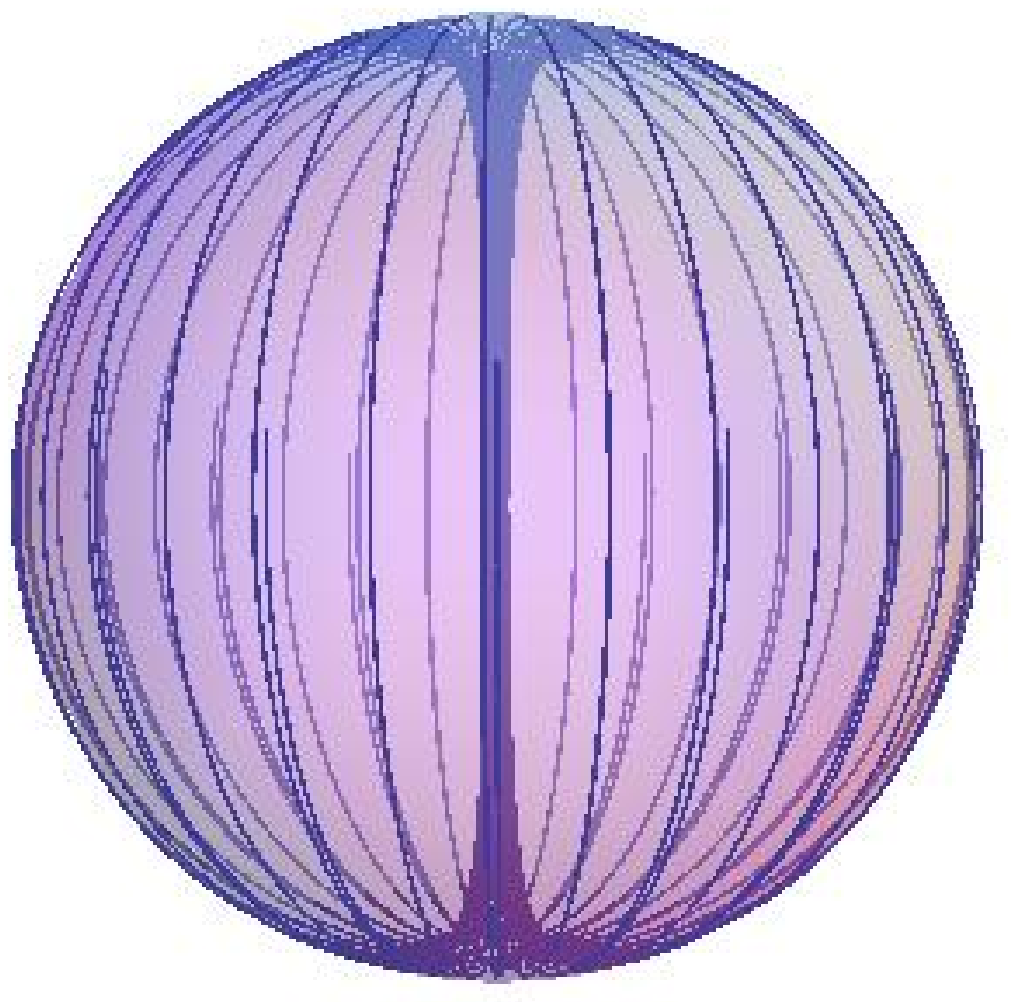}
\end{tabular}

&
\begin{tabular}{c}
% (b) \\
%\includegraphics[height=30mm]{sp11a449f.eps}
%\includegraphics[height=30mm]{1dtest.eps}
\includegraphics[height=30mm]{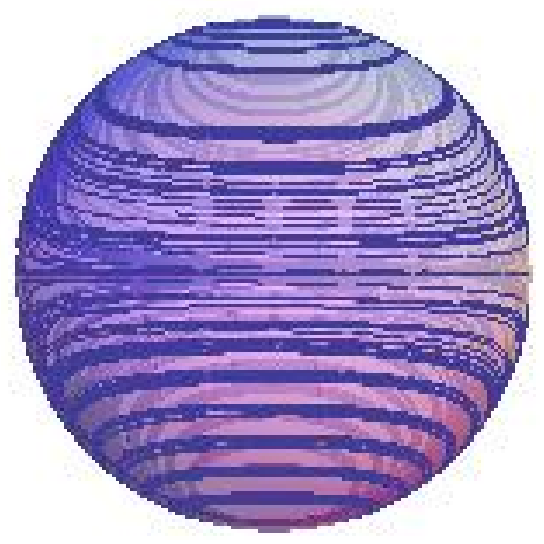}
\end{tabular}
\\

\begin{tabular}{c}
% (a) \\
\mbox{(2,0)}
\end{tabular}
&
\begin{tabular}{c}
% (a) \\
\mbox{(2,1)}
\end{tabular}
&
\begin{tabular}{c}
% (a) \\
\mbox{(2,2)}
\end{tabular}
\\
% (c) \\
\begin{tabular}{c}
\includegraphics[height=30mm]{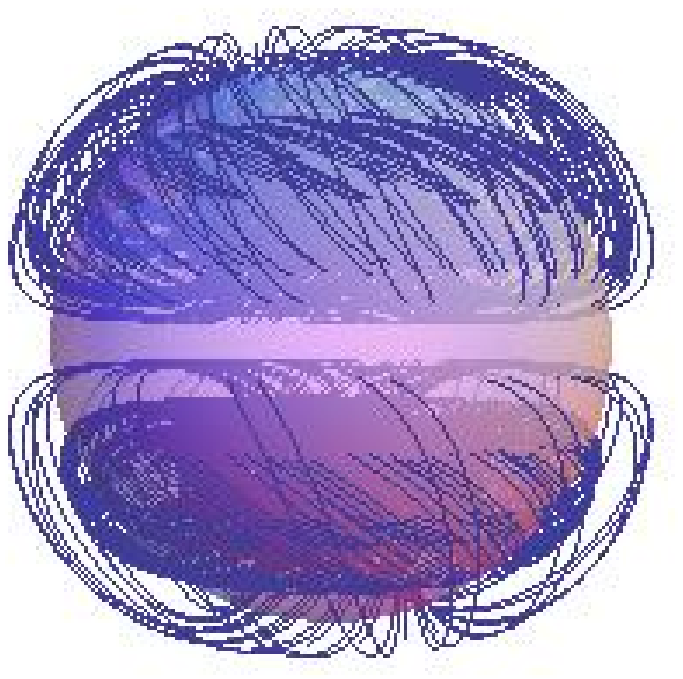}
\end{tabular}
&
\begin{tabular}{c}
% (d) \\
%\includegraphics[height=30mm]{sp21a449f.eps}
%\includegraphics[height=30mm]{1etest.eps}
\includegraphics[height=30mm]{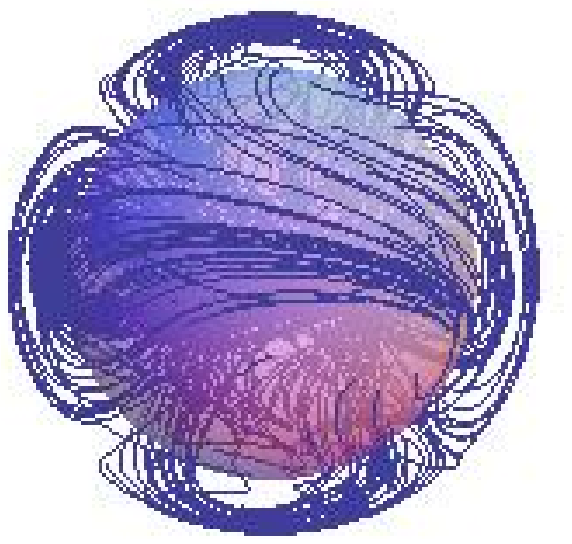}
\end{tabular}
&
\begin{tabular}{c}
% (d) \\
%\includegraphics[height=30mm]{sp22a449f.eps}
%\includegraphics[height=30mm]{1ftest.eps}
\includegraphics[height=30mm]{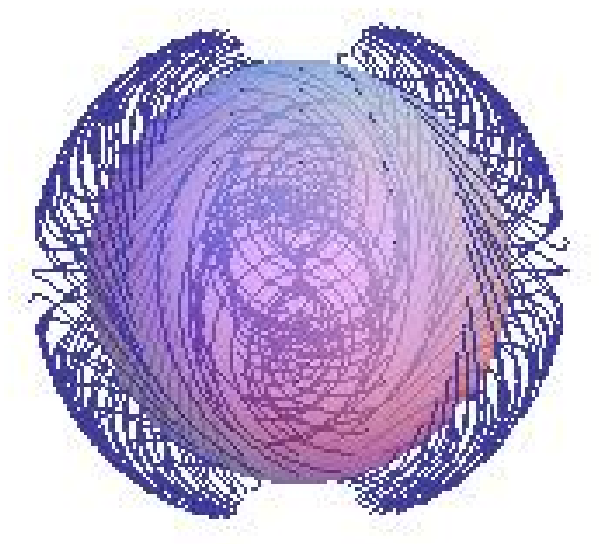}
\end{tabular}
\\
\begin{tabular}{c}
% (a) \\
\mbox{(3,0)}
\end{tabular}
&
\begin{tabular}{c}
% (a) \\
\mbox{(3,1)}
\end{tabular}
&
\begin{tabular}{c}
% (a) \\
\mbox{(3,2)}
\end{tabular}
&
\begin{tabular}{c}
% (a) \\
\mbox{(3,3)}
\end{tabular}
\\

% (e) \\
\begin{tabular}{c}
\includegraphics[height=30mm]{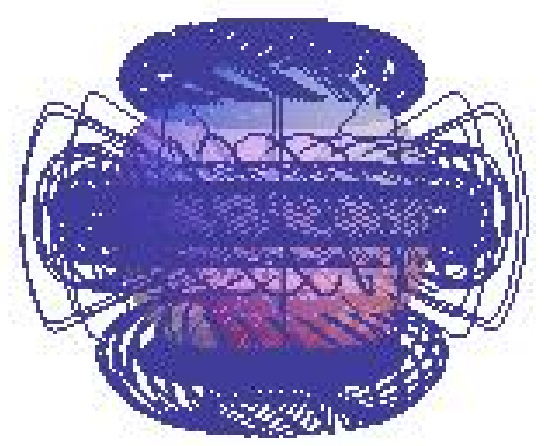}
\end{tabular}
&
\begin{tabular}{c}
% (d) \\
%\includegraphics[height=30mm]{1gtest.eps}
\includegraphics[height=30mm]{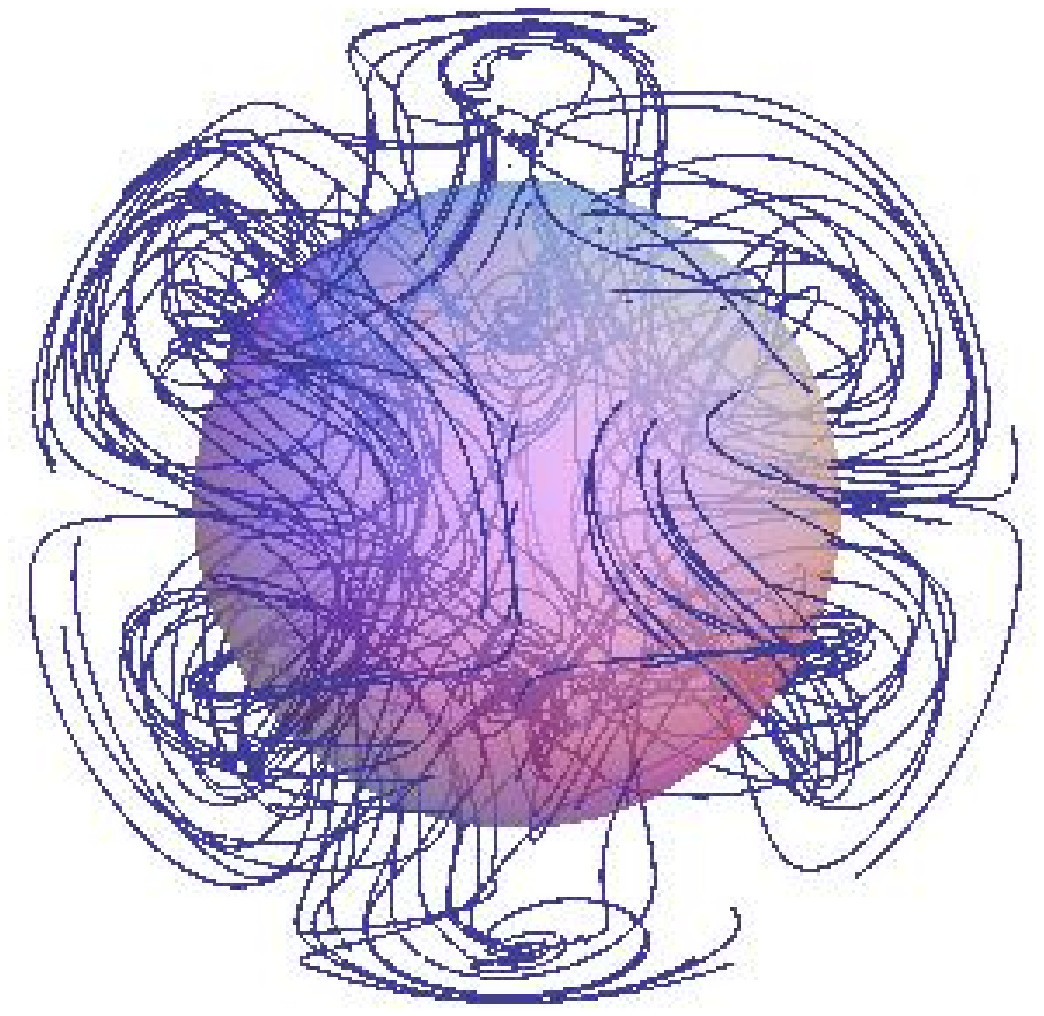}
\end{tabular}
&
\begin{tabular}{c}
% (d) \\
%\includegraphics[height=30mm]{sp32a449f.eps}
%\includegraphics[height=30mm]{1htest.eps}
\includegraphics[height=30mm]{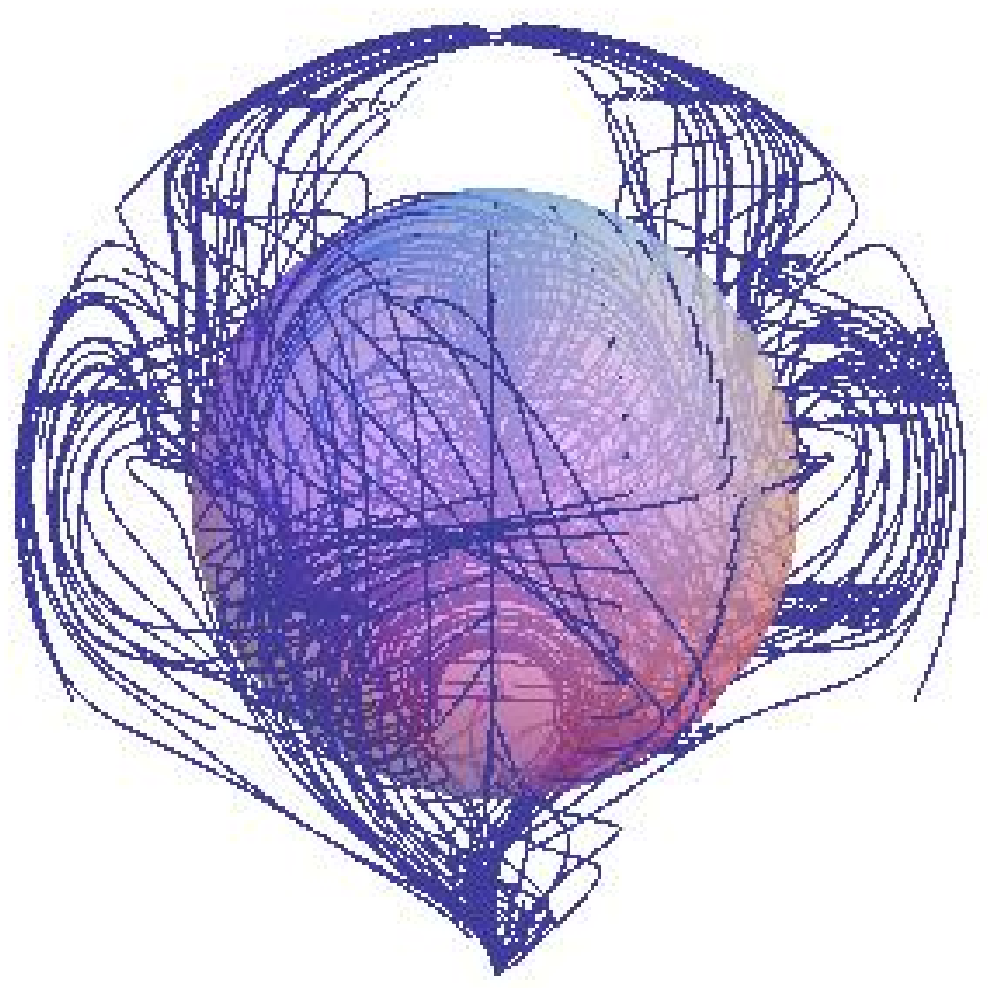}
\end{tabular}
&
\begin{tabular}{c}
% (d) \\
%\includegraphics[height=30mm]{1itest.eps}
\includegraphics[height=30mm]{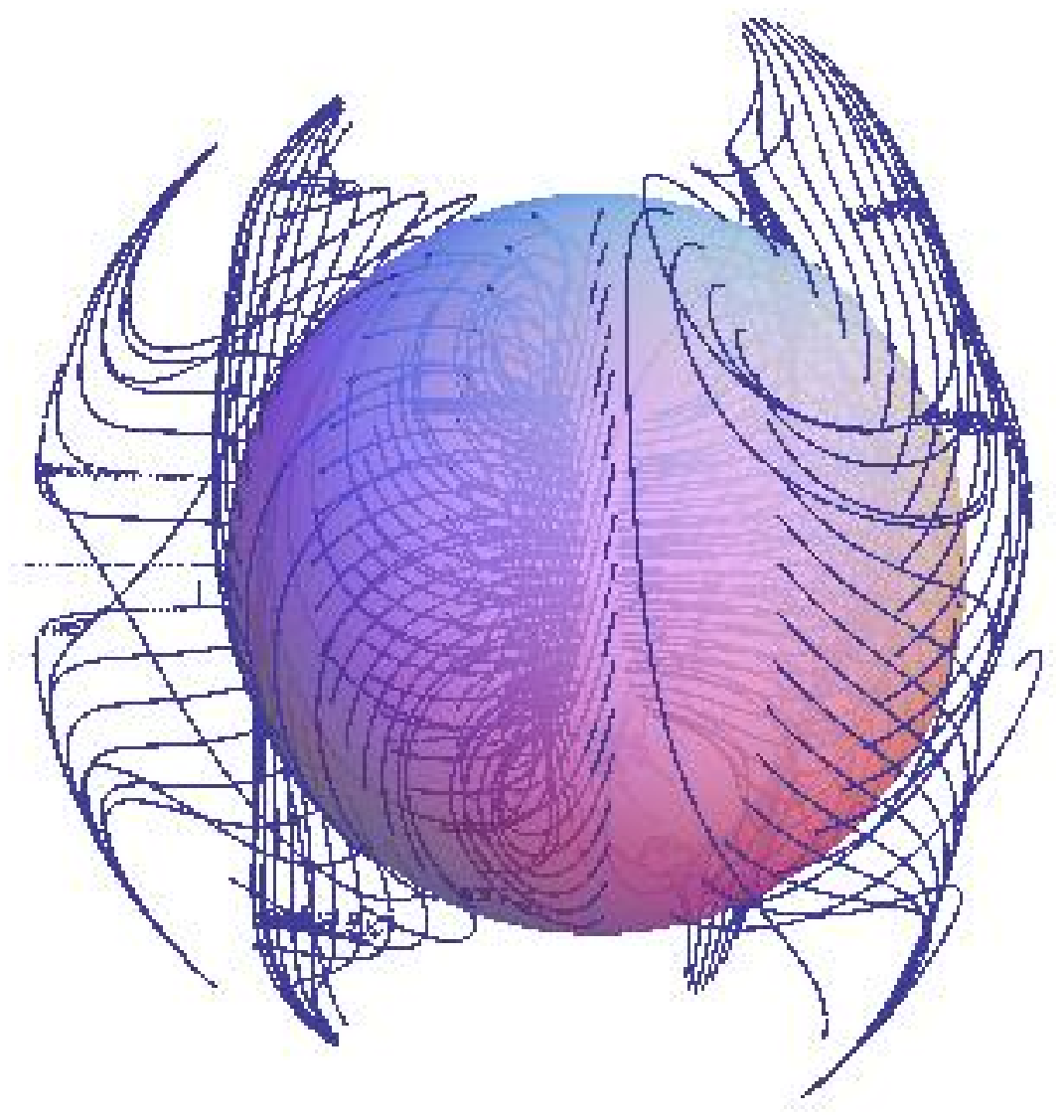}
\end{tabular}

\end{tabular}

\caption{Magnetic field lines for the force-free modes $1\leqslant n
\leqslant 3$, $0\leqslant m \leqslant n$, with the magnetic foot
points (the starting points for the field line integration) initialised at $ar_s=4.49$,
$\arccos(\theta)=(-8/9,-7/9,...,7/9,8/9)$,
$\phi=(0,\pi/8,\pi/4,...,15\pi/8)$. Each panel is labelled by
$(n,m)$. The north-south axis lies vertical in the plane of the
page.}
\end{figure*}
\end{center}

\subsection{Helicity}

The magnetic helicity within a volume $V$ is defined as

\be H=\int_V dV {\bf{A}}\cdot{\bf{B}},\ee where ${\bf{A}}$ is the
vector potential. It has been argued theoretically \citep{T86}, and
observed in spheromak experiments
\citep{Jetal80,Tetal83,Jetal85,Betal86,Ketal86}, that $H$ remains
constant as a force-free field relaxes. More generally, if a nonzero
electric field ${\bf{E}}=-\nabla\phi -\partial{\bf{A}}/\partial t$
is included (where $\phi$ is the scalar potential), Faraday's law and the definitions of $H$ and ${\bf{A}}$ imply $dH/dt=d({\bf{A}}\cdot{\bf{B}})/dt=-2{\bf{E}}\cdot{\bf{B}}+\nabla\cdot(\phi{\bf{B}}+{\bf{A}}\times{\bf{E}})$, which one can then integrate over a volume $V$ to obtain \citep{BS05}

\be \frac{dH}{dt} = -2\int_V dV {\bf{E}}\cdot{\bf{B}} +
\int_{\partial V} d{\bf{S}}\cdot ({\bf{A}}\times{\bf{E}} + \phi
{\bf{B}}). \label{eq1}\ee The helicity is gauge-independent if
${\bf{B}}$ is strictly tangential on the bounding surface $\partial
V$, which is true for spheromaks but not for a neutron star, unless
we extend $\partial V$ to infinity. This makes sense only if
${\bf{B}}$ falls off fast enough, such that $H$ is finite. To
achieve this, the force-free field [which scales as
$|{\bf{B}}|\propto r^{-1}$ for large $r$, according to (3)--(5)]
must be matched to a potential field at $r=r_s$ (where $r_s$ is the
surface of the star), something that is difficult to do both in
practice and in principle (Sec. 6 further explains why) \citep{W06}. Consequently, in
this paper, we assume that the field is force-free everywhere while
integrating over the volume $r\leqslant r_s$ in (6) to keep $H$
finite. The results in Sections 4--5 must be interpreted with this
approximation in mind.

The reader should note that, for simplicity's sake, we implicitly assume that the field at $r>r_s$ is also described by Eqs. (3)--(5), not by a potential field. This means that, at infinity, the magnetic field does not approach a dipole field (which is physically unrealistic and contrary to pulsar observations). Ideally, we wish to use an external field which is more realistic, such as a dipole field or a dipole-dominated multipole field, but this raises the issue of boundary matching, which is nontrivial and is discussed further in Sec. 6.

\subsection{Energy}

In a force-free field with uniform $a$, we have $H=\int_V dV {\bf{A}}\cdot{\bf{B}} = (1/a) \int_V dV {\bf{B}}^2 = 2\mu_0 W/a$, where $W$ is the total magnetic energy and $\mu_0$ is the magnetic permeability of free space. This is true both
for each mode $(n,m)$ in (3)--(5) and for a linear superposition of
modes; it can be easily verified by direct integration that the cross-terms $\int_V dV {\bf{B}}_{(l_1,m_1)}\cdot{\bf{B}}_{(l_2,m_2)}$ with $l_1\neq l_2$ and/or $m_1\neq m_2$ all vanish because of the linear independence of the modes. For a particular value of $a$, higher mode numbers have
higher energies. Fig. 2 illustrates this trend for modes with
$1\leqslant n\leqslant 3$ by graphing $W$ against $a$ for fixed
$r_s$. The modes $(n,m) = (1,0)$ and $(1,1)$ have equal energies,
despite having different structures, because the energies in their
radial and tangential components are separately equal.

Taking the dot product of Faraday's law with ${\bf{B}}/(2\mu_0)$, then using ${\bf{E}}={\bf{J}}/\sigma-{\bf{v}}\times{\bf{B}}$ to eliminate ${\bf{E}}$, and finally integrating over a volume $V$, one obtains the following equation for energy evolution \citep{BS05},

\be \frac{dW}{dt} = -\int_V dV ({\bf{J}}\times{\bf{B}})\cdot
{\bf{v}} - \frac{1}{\sigma} \int_V dV {\bf{J}}^2  -
\frac{1}{\mu_0}\int_{\partial V} d{\bf{S}}\cdot
{\bf{E}}\times{\bf{B}}, \end{equation} where $\bf{v}$ is the
velocity of the charged fluid threaded by ${\bf{B}}$, $\sigma$ is
the conductivity, and ${\bf{J}}$ is the current density. For force-free fields, one can write $\mu_0{\bf{J}}=\nabla\times{\bf{B}}=a{\bf{B}}$. Thus, the first term on the right-hand side of Eq. (8) vanishes, giving ${\bf{E}}\times{\bf{B}}=({\bf{J}}/\sigma -{\bf{v}}\times{\bf{B}})\times{\bf{B}}=-({\bf{v}}\times{\bf{B}})\times{\bf{B}}$ and hence

\be \frac{dW}{dt} = -\frac{a^2}{\mu_0^2\sigma}\int_V{\bf{B}}^2 dV +
\frac{1}{\mu_0}\int_V d{\bf{S}}\cdot
[({\bf{v}}\times{\bf{B}})\times{\bf{B}}].\ee

\begin{figure}
%\centerline{\epsfxsize=12cm\epsfbox{epplot.eps}}
\centerline{\epsfxsize=12cm\epsfbox{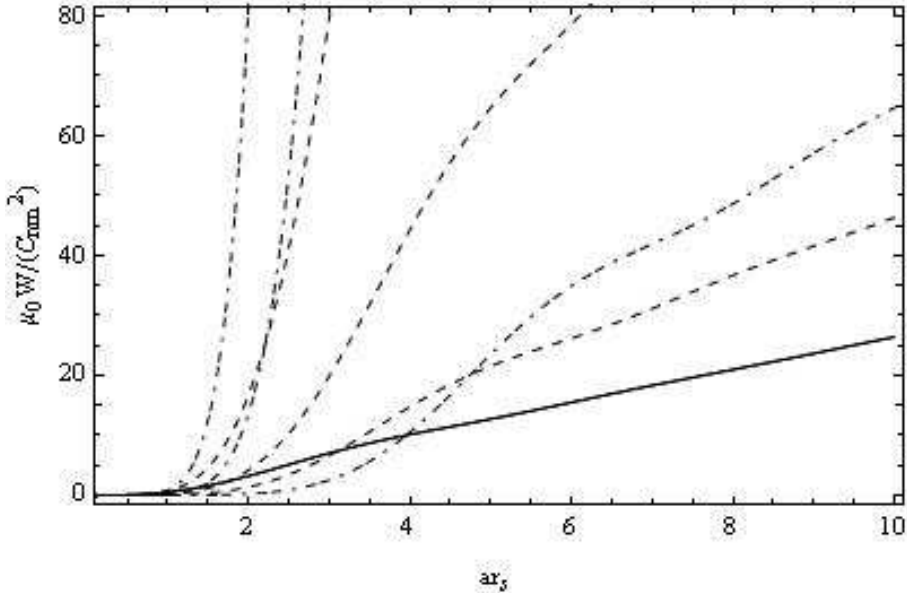}}
\caption{Magnetic energy $W_{nm}$ for the force-free mode $(n,m)$ in
(3)--(5) (in units of $C_{nm}^2/\mu_0$ as a function of $ar_s$,
where $a$ is the force-free parameter and $r_s$ is the radius of the
star. Solid curves: $n=1$; dashed curves: $n=2$; dashed-dotted
curves: $n=3$. For each value of $n$, the lowest curve corresponds
to the axisymmetric ($m=0$) and the highest curve corresponds to the
$m=n$ mode. Note that $W_{10}=W_{11}$ and $W_{30}=W_{31}$.}
\end{figure}

\section{Pre-diffusion Evolution}

The numerical experiments of Braithwaite and collaborators
\citep{BN06,BS04,BS06} proceeded in two stages. In the first stage,
a random magnetic field threading a stationary ball of plasma is
loaded into a three-dimensional MHD solver with both numerical and physical resistivities. We study the ideal-MHD evolution in this section. In the
second stage, the collisional resistivity is switched on. We study
the resistive evolution in Section 5. Note that we do not mean to imply here that the only diffusion process is Spitzer resistivity [another possibility is, for example, ambipolar diffusion \citep{GR92}]. Our results apply regardless of the actual diffusion mechanism, provided it is isotropic (i.e., $\sigma$ is a scalar).

\subsection{Initial conditions: random field}

Braithwaite and Nordlund (2006) initialised their simulation with a
random field, whose maximum wavenumber in a Fourier decomposition is
set by the grid scale ($\sim 10^2 R_*^{-1}$). We simulate the
initial state as a linear combination of the modes $1\leqslant n
\leqslant 3$, $0\leqslant m \leqslant n$, in which the modes are
equally weighted, i.e. all constants $C_{nm}$ are set to be equal.
We draw the field lines of the initial state in Fig. 4a. Note that
the maximum wavenumber is $k_{\mathrm{max}}\approx 4.2\pi$ (for
$n=m=3$). One can readily increase $k_{\textrm{max}}$ by increasing
$n$ to match Braithwaite and Nordlund's (2006) experiment, but there
is no need; the conserved global quantities are approximated to
$\sim 10$ \% with $n\leqslant 3$.

Note that the total energy $W_{\mathrm{T}}=\sum_{n=1}^{3}
\sum_{m=0}^{n} W_{nm}$ for our initial state is minimised at
$a=7.0$.

\subsection{Relaxation into equipartition}

We now test the hypothesis that the force-free magnetic field
evolves to energy equipartition. That is, we suppose that the
relative weightings $(C_{nm})$ of the modes adjust until each mode
contributes equally to the total energy,
$W_{\mathrm{T}}=N(N+1)W_{10}$ (for $1\leqslant n \leqslant N$,
$0\leqslant m\leqslant n$). In ideal MHD, this cannot happen
resistively. We postulate instead that it happens in the simulations
because the random initial perturbations $\delta {\bf{B}}$ (and
concomitant fluid velocity perturbations $\delta {\bf{v}}$) conspire
to produce a nonzero mean electromotive field $\langle
\delta{\bf{v}}\times \delta {\bf{B}} \rangle \neq 0$ as they
``unwind", just like the $\alpha$ effect in mean-field MHD
\citep{B93}.

A force-free magnetic field relaxing via the $\alpha$ effect seeks
out a state which minimizes its helicity at a nonzero value, as
observed in spheromaks \citep{T86}. We find $\partial H/\partial
a=0$ at $ar_s=4.49$ (where $r_s$ is the radius of the model neutron
star), the same conclusion reached by Taylor (1986) (see Fig. 3).
This result applies to any combination of modes in equipartition,
because $W_{\mathrm{T}}$ is proportional to $W_{10}$, and the energy
of the $(1,0)$ mode is minimized at $ar_s=4.49$. Hence, Taylor's
minimum eigenvalue for axisymmetric spheromak eigenfunctions applies
just as well to nonaxisymmetric eigenfunctions if the modes tend
towards equipartition. We plot the helicity and its derivative for
the composite equipartitioned state as functions of $a$ in Fig. 3;
the local minimum at $a=4.49 r_s^{-1}$ is apparent. The helicity is
calculated over the stellar volume, not out to infinity. This choice
is bound up in the physics of the boundary conditions, which is
discussed further at the end of this section and in Section 6.

\begin{figure}
%\centerline{\epsfxsize=12cm\epsfbox{htadhdatcom.eps}}
\centerline{\epsfxsize=12cm\epsfbox{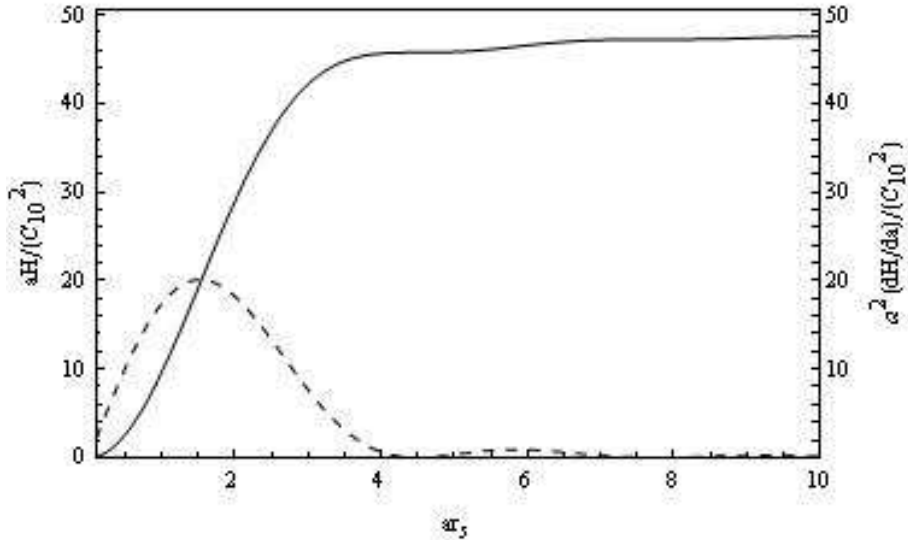}}
\caption{Helicity [solid curve, in units of $(C_{10})^2/a$] and its
derivative with respect to $a$ [dashed curve, in units of
$(C_{10})^2/a^2$] versus $ar_s$, for the composite equipartitioned
state with $n=1,2,3$ and $0\leqslant m\leqslant n$.}
\end{figure}

The total helicity $H_{\mathrm{T}}$ is related to the total energy
by $H_{\mathrm{T}}=2\mu_0W_{\mathrm{T}}/a$. Therefore, if the field
is to change from a random configuration (where all $C_{nm}$ are
equal) to an equipartitioned one (where all modes contribute equal
energy) while keeping both helicity and energy constant, $a$ must
remain constant as well. We designate this final state as State A.
If, on the other hand, the field is allowed to convert its energy
into, say, kinetic energy (e.g., random fluid motions in the MHD
simulation), $a$ can change to minimize $W_{\mathrm{T}}$
independently. We designate the corresponding final state as State
B. We present the values of the mode coefficients $C_{nm}$ achieved
through these two scenarios in Table 1. Note that, for all four
states, all constants $C_{nm}$ increase from their initial values
except for the (3,2) and (3,3) modes. Although $C_{nm}$ takes higher
values in State B than in State A, State B actually has a lower
magnetic energy, due to the decrease in $a$.

\begin{table*}
 \centering
 \begin{minipage}{170mm}
  \caption{Parameters and constants of the initial and final states of the force-free magnetic field, under four evolutionary scenarios without diffusion. The force-free parameter $a$ is measured in units of $r_s^{-1}$. The constants $C_{nm}$ are measured in units of $C_{10}$ at the start of the pre-diffusion phase, denoted by $C$.}
  \begin{tabular}{@{}lcccccccccc@{}}
  \hline
      Scenario&$a r_s$& $C_{10}/C$&$C_{11}/C$&$C_{20}/C$&$C_{21}/C$&$C_{22}/C$&$C_{30}/C$&$C_{31}/C$&$C_{32}/C$&$C_{33}/C$ \\
\hline Initial state&7.0&1.0&1.0&1.0&1.0&1.0&1.0&1.0&1.0&1.0 \\
State A: equipartition,&7.0&11&11&8.1&4.7&2.3&7.0&7.0&0.90&0.37\\
\quad $H_{\mathrm{T}}$, $W_{\mathrm{T}}$, and $a$ constant&&&&&&&&&\\
State B:
equipartition,&4.5&11&11&8.4&4.8&2.4&8.8&8.8&1.1&0.47\\
\quad $H_{\mathrm{T}}$ constant&&&&&&&&&\\
State C: mean-field MHD,&7.0&20&14&8.0&4.1&1.6&5.6&5.3&0.60&0.21\\
\quad $H_{\mathrm{T}}$, $W_{\mathrm{T}}$, and $a$ constant&&&&&&&&&\\
State D: mean-field MHD&4.5&21&15&8.1&4.2&1.7&5.7&5.4&0.61&0.21\\
\quad $H_{\mathrm{T}}$ constant&&&&&&&&&\\

\hline
\end{tabular}
\end{minipage}
\end{table*}

%&$1.45\times 10^{10}$ & $1.45\times 10^{10}$ &$9.32\times 10^9$

\begin{center}
\begin{figure*}
%\begin{tabular}{ccc}
%\begin{tabular}{c}
%\end{tabular}
%&
%\begin{tabular}{c}
\mbox{(a)}
%\end{tabular}
%&
%\begin{tabular}{c}
%\end{tabular}
\\
%\begin{tabular}{c}
%\end{tabular}
%&
%\begin{tabular}{c}
% (a) \\
%\includegraphics[height=50mm]{a7initn123allc1front.eps}
%\includegraphics[height=70mm]{a7test.eps}
\includegraphics[height=75mm]{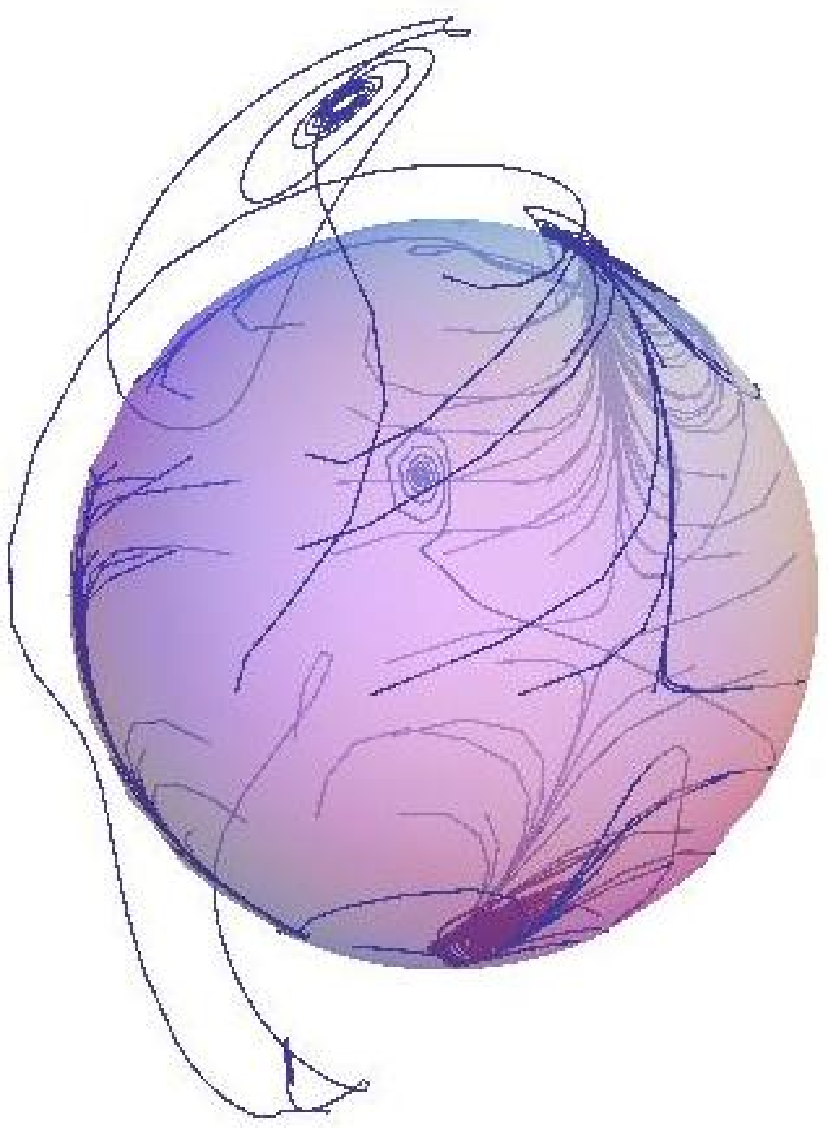}
%\includegraphics[width=2.0in, angle=-90]{Fig/monte_8.epsi}
%\end{tabular}
%&
%\begin{tabular}{c}
%\end{tabular}
\\
%\begin{tabular}{c}
\mbox{(b)}
%\end{tabular}
%&
%\begin{tabular}{c}
%\end{tabular}
%&
\mbox{} \mbox{} \mbox{}\mbox{} \mbox{} \mbox{}\mbox{} \mbox{}
\mbox{}\mbox{} \mbox{} \mbox{}\mbox{} \mbox{} \mbox{}\mbox{} \mbox{}
\mbox{}\mbox{} \mbox{} \mbox{}\mbox{} \mbox{} \mbox{}\mbox{} \mbox{}
\mbox{}\mbox{} \mbox{} \mbox{}\mbox{} \mbox{} \mbox{}\mbox{} \mbox{}
\mbox{}\mbox{} \mbox{} \mbox{}\mbox{} \mbox{} \mbox{}\mbox{} \mbox{}
\mbox{}\mbox{} \mbox{} \mbox{}\mbox{}\mbox{} \mbox{} \mbox{}\mbox{}
\mbox{} \mbox{}\mbox{} \mbox{} \mbox{}\mbox{} \mbox{} \mbox{}\mbox{}
\mbox{} \mbox{}\mbox{} \mbox{} \mbox{}\mbox{} \mbox{} \mbox{}\mbox{}
\mbox{}\mbox{} \mbox{} \mbox{}\mbox{} \mbox{} \mbox{}\mbox{} \mbox{}
\mbox{}\mbox{} \mbox{}
%\begin{tabular}{c}
\mbox{(c)}
%\end{tabular}
\\
% (c) \\
%\begin{tabular}{c}
%\includegraphics[height=20mm, angle=-90,width=0.49\linewidth ]{Fig/monte_9.epsi}
%\includegraphics[height=73mm]{a7equipn123m0123c11052front.eps}
\includegraphics[height=73mm]{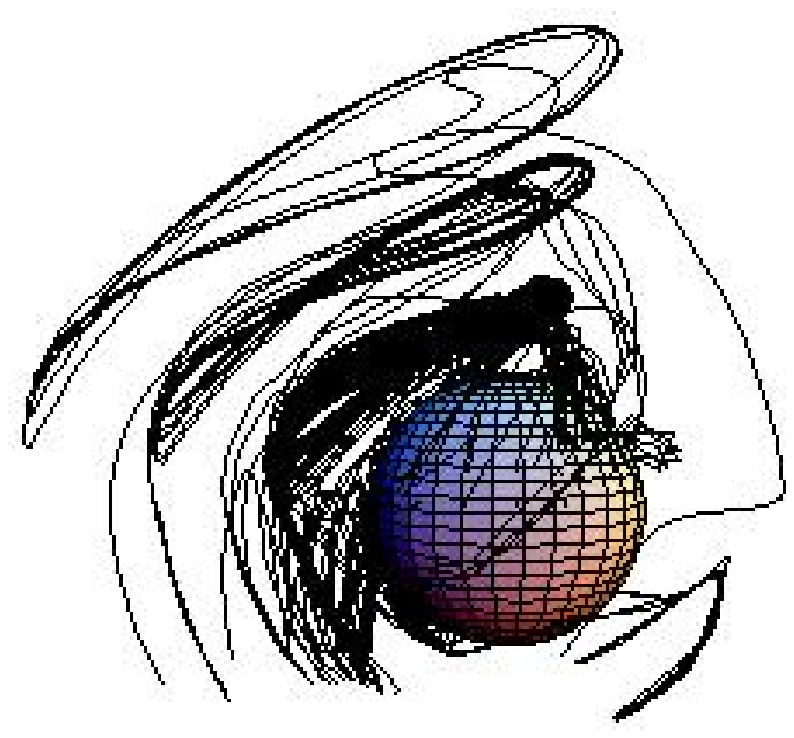}
%\end{tabular}
%&
%\begin{tabular}{c}
%\end{tabular}
%&
%\begin{tabular}{c}
% (d) \\
%\includegraphics[height=73mm]{a449equipn123c11067676front.eps}
\includegraphics[height=73mm]{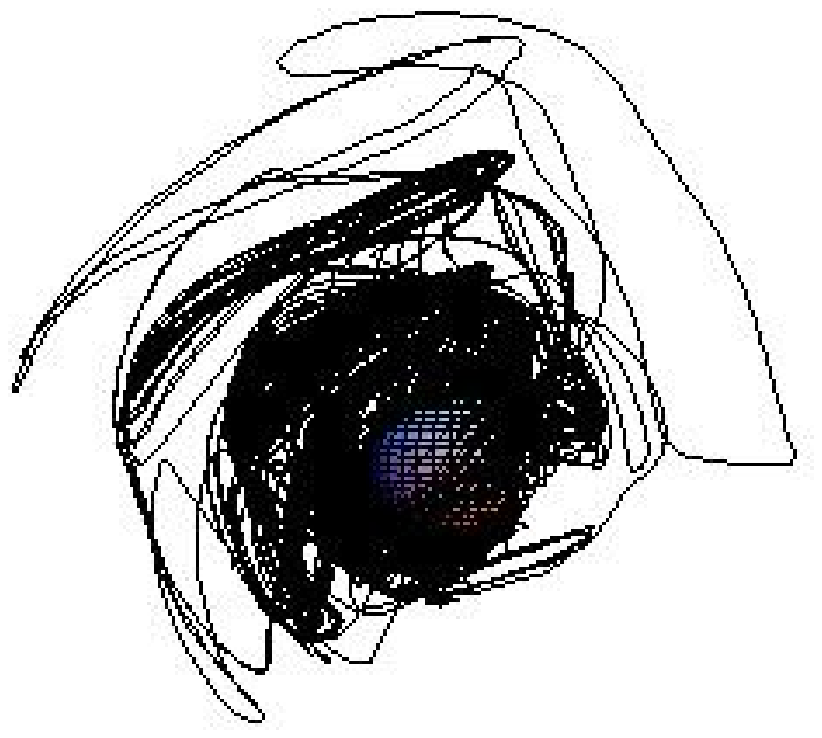}
%\includegraphics[width=2.0in, angle=-90]{Fig/monte_3.epsi}
%\end{tabular}
%\\

%\end{tabular}

\caption{Magnetic field configurations produced by the evolutionary
hypotheses in Section 4.2. (a) Initial random field, with $C_{nm}=1$
for all $1\leqslant n \leqslant 3$, $0\leqslant m\leqslant n$, and
$ar_s=7.0$. (b) Final force-free field in State A, with $C_{10}=11$,
$C_{11}=11$, $C_{20}=8.1$, $C_{21}=4.7$, $C_{22}=2.3$, $C_{30}=7.0$,
$C_{31}=7.0$, $C_{32}=0.90$, $C_{33}=0.37$, and $ar_s=7.0$. (c)
Final force-free field in State B, with $C_{10}=11$, $C_{11}=11$,
$C_{20}=8.4$, $C_{21}=4.8$, $C_{22}=2.4$, $C_{30}=8.8$,
$C_{31}=8.8$, $C_{32}=1.1$, $C_{33}=0.47$, and $ar_s=4.5$. The
shaded sphere is at a radius $r_s$.}

%\end{itemize}
\end{figure*}
\end{center}

\begin{center}
\begin{figure*}
\begin{tabular}{cc}

\begin{tabular}{c}
\mbox{(a)}
\end{tabular}
&
\begin{tabular}{c}
% (d) \\
\mbox{(b)}
\end{tabular}
\\
% (e) \\
\begin{tabular}{c}
\includegraphics[height=80mm]{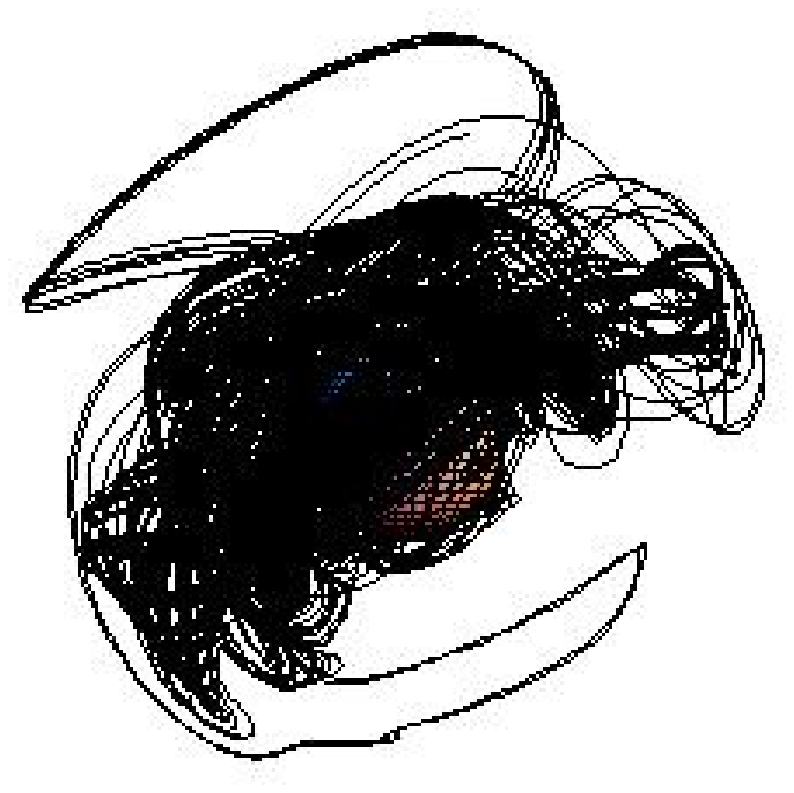}
\end{tabular}
&
\begin{tabular}{c}
% (d) \\
%\includegraphics[height=80mm]{a449n123m0123vii.eps}
\includegraphics[height=80mm]{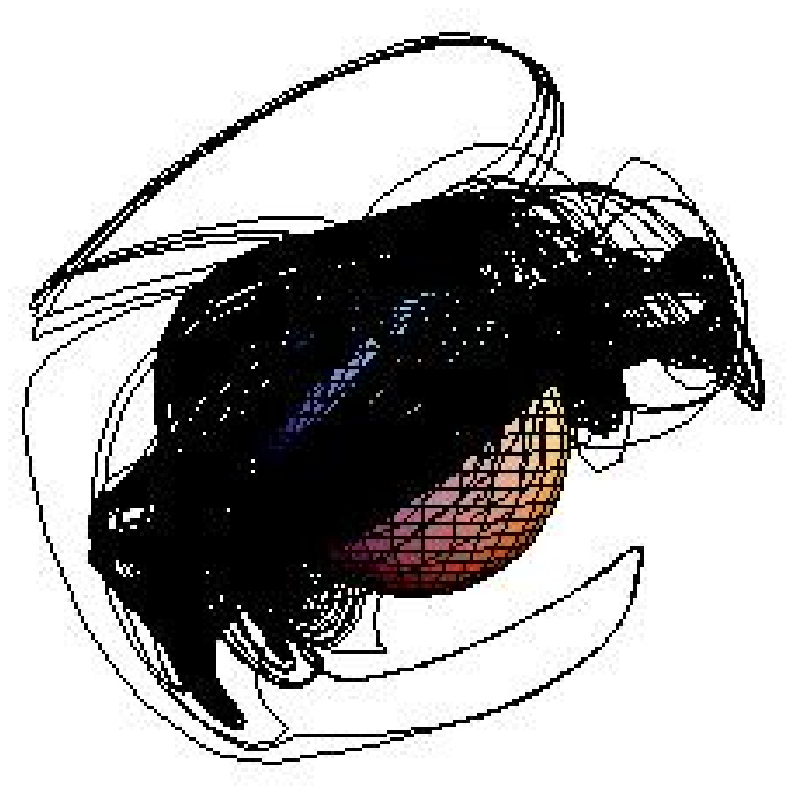}
\end{tabular}

\end{tabular}

\caption{Magnetic field configurations produced by the evolutionary
hypotheses in Section 4.3. (a) Final force-free field in State C,
with $C_{10}=20$, $C_{11}=14$, $C_{20}=8.0$, $C_{21}=4.1$,
$C_{22}=1.6$, $C_{30}=5.6$, $C_{31}=5.3$, $C_{32}=0.60$,
$C_{33}=0.21$, and $ar_s=7$. (b) Final force-free field in State D,
with $C_{10}=21$, $C_{11}=15$, $C_{20}=8.1$, $C_{21}=4.2$,
$C_{22}=1.7$, $C_{30}=5.7$, $C_{31}=5.4$, $C_{32}=0.61$,
$C_{33}=0.21$, and $ar_s=4.5$. The shaded sphere is at a radius
$r_s$.}
\end{figure*}
\end{center}

We plot magnetic field lines of the initial state, State A, and
State B in Figs. 4a, 4b, and 4c respectively. The two final states
are similar in shape, with State B having a larger radius
$r_{\mathrm{M}}$, defined as \citep{BN06}

\be r_{\mathrm{M}} = \left(\frac{\int dV B^2r^2}{\int dV
B^2}\right)^{1/2}.\ee As the star does not have an actual surface, we use $r_\mathrm{M}$ as a proxy for the characteristic spatial extent of the field. Specifically, we find $r_{\mathrm{M}}/r_s=
0.71$, 0.85, and 1.0 for the initial state, State A, and State B
respectively. Fig. 4 is to be compared with the bottom left panel of
Fig. 6 in Braithwaite and Nordlund (2006), which depicts the state
achieved by the ideal-MHD simulation after 9 Alfv\'{e}n crossing
times of nondiffusive evolution. Qualitatively, the results are
similar. Note how field lines form toroidal loops around the star in
both the numerical and analytical calculations. The toroidal loops
are wound closer to the star in Fig. 4c than in Fig. 4b, indicating
that State B resembles Braithwaite and Nordlund (2006) more closely
than State A. Also note that the ``surface" of the star is located at $r=r_s$ (where we also initialise the field lines), unlike in Fig. 6 of Braithwaite and Nordlund (2006), where the sphere is at $r=0.3r_s$. Our model star does not have a hard boundary, and we do not match the ``interior" force-free field to an ``exterior" potential field at $r=r_s$.

Imposing a boundary condition at $r=r_s$, such as a conducting wall
(more appropriate for laboratory plasmas) \citep{S90}, balance
between external plasma pressure and internal magnetic pressure
\citep{Z96,Vetal03}, and force balance with an external magnetic
field \citep{LB03}, affects the relaxation path and final state
\citep{T86,BE04}. The problem of matching a force-free field to an
external field is nontrivial \citep{BN07}, requiring $a$ to vary
(dis)continuously along the interface (plus the existence of
non-force-free regions in the star). This matter is discussed in
more detail in Section 6. Another potential issue is that, while the conductivity certainly damps the higher modes more quickly in a realistic field, as in the simulations, this is not true for the special case of force-free fields, where $\nabla^2 {\bf{B}} = -a^2{\bf{B}}$ for all $(l,m)$. This is another argument against the realism of force-free fields.

\subsection{Blackman and Field (2004) mean-field dynamo}

In the previous section, we do not advance a theory for how the
parameter $a$ and the constants $C_{nm}$ change with time. We simply
postulate an end state which fulfills energy equipartition and which
is linked to the initial state by globally conserved quantities. In
this section, we model the evolution more closely by applying a
model developed by Blackman and Field (2004). This model is intended
by the authors to be applicable to both astrophysical phenomena
(such as stellar coronae) and laboratory plasma experiments.
Blackman \& Field (2004) explicitly sidestepped discussion of
physical boundary conditions (particularly current closure) by
utilizing periodic boundary conditions instead.

The model, rooted in the principles of mean-field MHD, gives a
recipe for the coupled evolution of the spatially-averaged helicity
on two length scales (one large, one small), as well as the
spatially-averaged fluid vorticity and electromotive field. Upon
splitting the magnetic field, vector potential, and current density
into (slowly varying) large-scale and (turbulent) small-scale
components, the volume-averaged induction equation and Faraday's law
imply

\be \partial_t
H_1=2\langle{\bf{\overline{\varepsilon}}}\cdot{\bf{\overline{B}}}\rangle-2\eta\langle{\bf{\overline{B}}}\cdot{\bf{\overline{J}}}\rangle\ee
and

\be \partial_t
H_2=2\langle{\bf{\overline{\varepsilon}}}\cdot{\bf{\overline{B}}}\rangle-2\eta\langle{\bf{b}}\cdot{\bf{j}}\rangle.\ee
In (12)--(13), $\eta$ is the magnetic diffusivity, $H_{1}$ and $H_2$
are the large- and small-scale magnetic helicities respectively,
${\bf{\overline{B}}}$ and ${\bf{b}}$ are the large- and small-scale
magnetic fields, ${\bf{\overline{J}}}$ and ${\bf{j}}$ are the large-
and small-scale current densities, and
${\bf{\overline{\varepsilon}}}=-{\bf{\overline{E}}}+\eta\langle{\bf{\overline{B}}}\cdot{\bf{\overline{J}}}\rangle$
is the turbulent electromotive field. By utilising Amp\`{e}re's Law,
Faraday's Law, and the incompressible Navier-Stokes equation,
Blackman and Field (2004) derived the following closed set of
evolution equations for the spatially averaged helicities ($H_1$ and
$H_2$), vorticity $H_V=\langle
{\bf{v}}\cdot\nabla\times{\bf{v}}\rangle$, electromotive field $Q$,
and energy $\epsilon$:

\be \frac{\partial h_1}{\partial \tau} =
-2Qh_1^{1/2}(k_1/k_2)^{1/2}-2h_1(k_1/k_2)^2/R_M,\ee \be
\frac{\partial h_2}{\partial \tau} =
2Qh_1^{1/2}(k_1/k_2)^{1/2}-2h_2/R_M,\ee \be \frac{\partial
h_v}{\partial
\tau}=-2(1-k_1/k_2)(k_1/k_2)^{1/2}Qh_1^{1/2}-2h_v/R_V,\ee \be
\frac{\partial Q}{\partial
\tau}=-(k_1/k_2)^{1/2}h_1^{1/2}(1/3)(h_2-h_v)+(k_1/k_2)^{3/2}h_1^{1/2}\epsilon/3-\epsilon^{1/2}fQ,\ee
\be \frac{\partial \epsilon}{\partial
\tau}=-2(1-k_1/k_2)(k_1/k_2)^{1/2}Qh_1^{1/2}-2\epsilon/R_V.\ee In
(14)--(18), we define $k_1$ and $k_2$ as the characteristic large-
and small-scale wavenumbers, $R_V=H_2(0)^{1/2}/\nu k_2^{1/2}$ (where
$\nu$ is viscosity), $R_M=H_2(0)^{1/2}/\eta k_2^{1/2}$,
$h_v=H_V/k_2^2H_2(0)$, $h_1$ and $h_2$ are the large- and
small-scale helicities normalised to $H_2(0)$,
$Q=-{\bf{\overline{\varepsilon}}}_{\parallel}/k_2H_2(0)$ is the
component of the turbulent electromotive field parallel to
${\bf{\overline{B}}}$,
$\epsilon=\langle{\bf{v}}^2\rangle/k_2H_2(0)$, where $\bf{v}$ is the
turbulent fluid velocity, $\tau=tk_2^{3/2}H_2(0)^{1/2}$, and $f\sim
1$ is a dimensionless constant which parametrizes the (unspecified)
microphysical dissipation \citep{BF04}. The notation `$(0)$' denotes
an initial value.

Let us now look at what happens in the steady state. From (14) and
(15), the steady-state helicities satisfy

\be h_2=(k_1/k_2)^2 h_1.\ee For our force-free modes (3)--(5), we
can write $k_{mn}\propto (n^2+m^2)^{1/2}$. Suppose that the two-mode
result (19) can be generalized to apply to all the modes in a linear
combination. We designate the final relaxed state with $C_{nm}$
related through (19) as State C if both helicity and magnetic energy
are kept constant throughout relaxation, and as State D if only the
helicity is kept constant. Magnetic field lines for States C and D
are shown in Figs. 5a and 5b respectively. These states \emph{do
not} obey the equipartition relation. Note that States C and D are
very similar (see also Table 1), while A and B are less so,
indicating a weaker dependence of $H_{\mathrm{T}}$ on $a$. The
values of $r_{\mathrm{M}}/r_s$ are 0.77 and 0.78 for States C and D
respectively. As evident from Table 1, all $C_{nm}$ in States C and
D are smaller than in States A and B, except for $C_{1m}$. In fact,
the axisymmetric (1,0) mode dominates in States C and D, being
approximately 20 times greater than in the initial state, giving
these states a more toroidal shape (see Figs. 5a and 5b).

To quantify further the `shape' of the field, we compute the
poloidal flux $\Phi_{\theta}$ and toroidal flux $\Phi_{\phi}$ from

\be \Phi_{\theta} = \int_{0}^{2\pi} d\phi\int_{0}^{r_s} dr
\phantom{+}rB_\theta(r,\pi/2,\phi) \ee \be \Phi_\phi =
\frac{1}{2}\int_{0}^{\pi}d\theta \int_0^{r_s} dr
\phantom{+}rB_{\phi}(r,\theta,0) + \frac{1}{2}\int_{0}^{\pi}d\theta
\int_0^{r_s} dr \phantom{+}rB_{\phi}(r,\theta,\pi).\ee The results
are listed in Table 2. In all four final states, $\Phi_\theta$ and
$\Phi_\phi$ are comparable. This is a natural property of the linked
toroidal-poloidal field structure necessary for long-term MHD
stability. Again, we see that nonaxisymmetric force-free modes play
a key role, this time in ensuring long-term stability.

\begin{table*}
 \centering
 \begin{minipage}{170mm}
  \caption{Poloidal flux, toroidal flux, and ratio of poloidal flux to toroidal flux, for the random initial state and States A--D. The fluxes are expressed in units of $C r_s^{-2}$, where $C$ is the initial value of the coefficient $C_{10}$.}
  \begin{tabular}{@{}lccc@{}}
  \hline
     State &Poloidal flux& Toroidal flux&Poloidal flux/toroidal flux \\
\hline Random state &1.3&2.6&0.50 \\
State A &14&20&0.70\\
State B &34&31&1.1\\
State C &27&22&1.2\\
State D &22&47&0.47\\

\hline
\end{tabular}
\end{minipage}
\end{table*}

\section{Diffusive Evolution}

In the second stage of the numerical experiments described by
Braithwaite and Nordlund (2006), the resistivity is switched on and
the magnetic field obtained at the end of the (ideal-MHD) first
stage is allowed to relax by Ohmic diffusion \citep{BS06}. What are
the implications for our analytic model?

If the field remains in a linear superposition of quasistatic
force-free modes,  the linear independence of the modes ensures that
they all decay at the same rate, proportional to the electrical
conductivity $\sigma$:

\be dC_{nm}/dt=(a^2/\mu_0\sigma)C_{nm}.\ee If the force-free
parameter $a$ also remains constant throughout the diffusion
process, it is evident that the structure of field lines, e.g. the
quantity $r_{\mathrm{M}}$ and poloidal-to-toroidal flux ratio, also
remain unchanged. This is contrary to the results of Braithwaite and
Nordlund, who found that the field becomes more poloidal and
dipole-like as it diffuses outwards \citep{BN06}. By implication,
the field does not remain force free (with constant $a$) in the
resistive stage of the simulations. We note that an extra term is added to the magnetic field diffusion equation if $a$ is nonuniform, since the term $\eta \nabla^2 {\bf{B}}$ now becomes $-a^2\eta{\bf{B}}-\eta(\nabla a)\times {\bf{B}}$, which affects the shape of the field as it diffuses, depending on $\nabla a$.

However, if we allow the energy to vary freely and even allow some
loss of helicity \citep{BS06}, some of the magnetic energy at the
start of the resistive stage is converted into fluid kinetic energy
$\epsilon$, inducing a turbulent dynamo which regenerates the field
(or, at least, exchanges helicity between small and large scales).
Following Blackman and Field (2004), we consider two force-free
modes with sufficiently different characteristic wavenumber $k$
[modes (1,1) and (3,1), for example] and let them evolve from
equipartition as per (14)--(18). The results are plotted in Fig. 6
and reveal some general trends. First of all, we find that the
turbulent vorticity $h_v$, which starts at zero, increases briefly
then dies away, controlled by $R_V$ ($h_v$ falls away slower for
larger $R_V$, i.e. smaller viscosity). Secondly, the long-wavelength
mode increases at first but then decreases while the
short-wavelength mode decreases immediately (and faster). The rates
at which the modes decay are controlled by the quantity $R_M$; the
larger the magnetic diffusivity, the faster the decay. Thus, in the
long term, the lower $n$ and $m$ modes dominate, until all the
higher order and nonaxisymmetric modes vanish; a force-free field
which is a linear combination of several modes can decay into an
axisymmetric Taylor state, given enough time and a mediating
turbulent vorticity. We stress again, however, that this dynamo
picture is one of many possibilities, and it does not take into
account superfluid and stratification effects in a neutron star.

\begin{figure}
\centerline{\epsfxsize=12cm\epsfbox{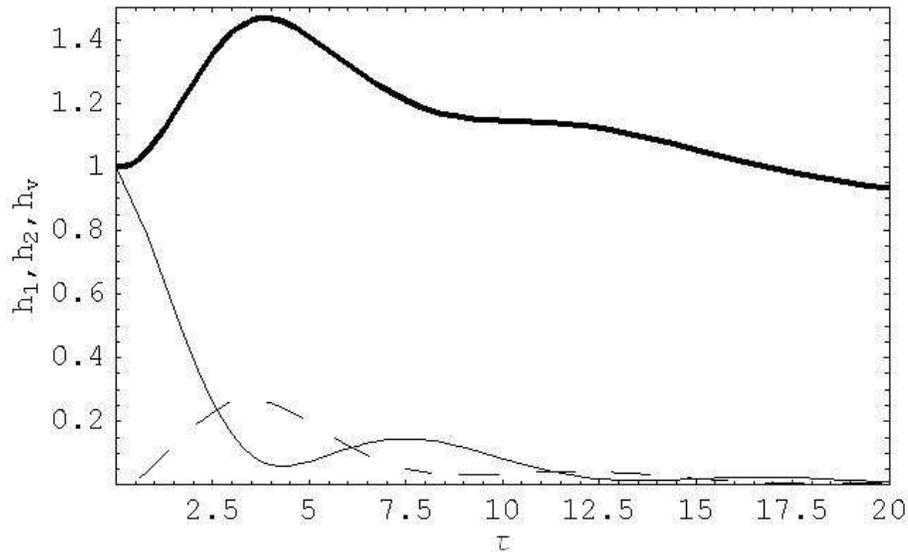}}
\caption{Dimensionless helicities of the (1,1) mode ($h_1$, thick
solid curve), (3,3) mode ($h_2$, thin solid curve), and vorticity
($h_v$, dashed curve) as functions of $\tau$. $h_1$ and $h_2$ are
initially equal (the modes are in equipartition) and $h_v$ is
initially zero. The dimensionless physical parameters $f$, $R_M$,
and $R_V$ are taken to be 1, 10, and 10 respectively.}
\end{figure}

\section{Discussion}

We show in this paper that an initial magnetic field which is a
linear combination of equally weighted, force-free modes can relax
without diffusion, preserving magnetic helicity, to an
equipartitioned state which possesses a linked poloidal-toroidal
geometry (with toroidal flux dominating), just like the end result
of the first (nondiffusive) stage of the numerical simulations by
Braithwaite and collaborators \citep{BN06,BS06}. Alternatively, a
turbulent, mean-field MHD dynamo model \citep{BF04} can be used to
prescribe the evolution of the force-free field, allowing helicity
to be exchanged between small and large scales through the agency of
a turbulent electromotive field (and growth of small-scale
hydrodynamic vorticity). Applying this particular model to our
initial state, we again obtain a final configuration which
qualitatively resembles the numerical results of Braithwaite and
Nordlund (2006), but with smaller root-mean-square size
$r_{\mathrm{M}}$. \emph{The inclusion of nonaxisymmetric force-free
modes is important to obtain good agreement.}

A force-free field decays uniformly when a uniform resistivity is
switched on, keeping the relative weighting of its modes (and hence
its three-dimensional structure) constant. This behaviour is not
observed in the numerical simulations, where instead the field
diffuses outwards over time into a dipolar shape \citep{BN06,BS06}.
If, however, a turbulent electromotive force is allowed to operate,
we show that the higher-order modes decay faster, and the magnetic
field tends overall towards the Taylor axisymmetric state (which is
dominated by toroidal flux at $ar_s=4.49$).

We conclude by speaking briefly about the vital issue of boundary
conditions at the stellar surface. In this paper, we assume (for the
sake of analytic simplicity) that the magnetic field is force-free
everywhere, because our main goal is to elucidate the central role
played by the nonaxisymmetric modes. In order to preserve finite
helicity and energy, we are forced to integrate (artificially)
${\bf{A}}\cdot {\bf{B}}$ and $B^2/2\mu_0$ over just the volume of
the star, rather than over all space. Hence the helicity and energy
we compute are semiquantitative approximations. It is natural to try
and correct this shortcoming by matching a force-free field in the
stellar interior to a current-free (potential) field in an
insulating atmosphere above the stellar surface \citep{BN07}. The
matching conditions are popularly taken as: (i) normal component of
${\bf{B}}$ is continuous at the surface (which uniquely determines
the external field) and (ii) the jump in the tangential component of
${\bf{B}}$ is proportional to the surface (skin) current (and indeed
uniquely determines the latter). However, this prescription hides a
subtle flaw: current continuity $(\nabla\cdot{\bf{J}}=0)$ is
violated at the surface, because it is impossible, in general, for
the skin current leaving (entering) a surface patch sideways to
balance the volume current $\mu_0^{-1}aB_r$ entering (leaving) the
patch from below. The root of this apparent paradox
concerning current closure can be traced back to the following
classic theorem of magnetostatics: a magnetic field which is force
free everywhere within a finite volume cannot match to an external
field which is potential everywhere \citep{P84}\footnote{The theorem
follows from the identity

\be \int dV {\bf{x}}\cdot ({\bf{J}}\times{\bf{B}}) =
\frac{1}{2\mu_0} \int dV |{\bf{B}}|^2 + \frac{1}{2\mu_0}\int
d{\bf{S}}\cdot[2({\bf{x}}\cdot{\bf{B}}){\bf{B}}-|{\bf{B}}|^2{\bf{x}}],\ee
if we consider a spherical surface of radius $R$ that fully encloses
the force-free region and let $R\rightarrow \infty$.}. [Physically, solutions with surface currents are not appealing, because conductivities drop off towards the surface of the star \citep{B08}. In this way, Broderick and Narayan's (2007) treatment is problematic; surface currents are an artifact arising out of the matching problem and are accompanied by an artificial surface ${\bf{J}}\times{\bf{B}}$ force.]. One way to
resolve the paradox is to allow currents to flow in the atmosphere
above the surface, e.g. through discrete structures like magnetic
arcades (cf. the Sun). Under these circumstances, the force-free
parameter $a({\bf{x}})$ must vary with position, and the matching
conditions at any interface reduce to continuity of (i) the normal
component of ${\bf{B}}({\bf{x}})$, and (ii) $a({\bf{x}})$ (over one
polarity of ${\bf{B}}$), which are solved iteratively \citep{W06}.
Loss of helicity through parts of the surface can occur
\citep{BS06}, although care must be taken to use appropriate
boundary conditions on the slow relaxation time-scale. Another way
to resolve the paradox is to create a non-force-free (i.e.,
stressed) region within the star, through which the currents can
close. Imposing boundary conditions at the stellar surface which
satisfy $\nabla\cdot{\bf{J}}=0$ in this scenario is a nontrivial
task.

Both resolutions of the matching paradox outlined above may carry
important implications for magnetar astrophysics. The first suggests
the likelihood of magnetic activity (e.g. reconnection between
magnetic arcades) in the magnetosphere, accompanied perhaps by radio
and X-ray bursts. The second suggests that magnetic stresses may
cause crust cracking in non-force-free regions. Note that neutron star magnetosphere models have popularly included surface currents in the past [e.g., Goldreich and Julian (1969), Ruderman (1975), Mestel \emph{et al.} (1985), and Arons (1993)]. This paper exposes
some of the limitations of analytic methods. A better way to study
such possibilities \emph{self-consistently} is to extend the
simulations in Braithwaite and Nordlund (2006) and elsewhere to
treat in more detail the mechanical behaviour of the stellar
interior.

\section*{Acknowledgments}

We thank Paul Cally for alerting us to the impossibility of matching
an internal force-free field to an external potential field. We also thank the reviewer for his/her invaluable input and constructive criticism, which have improved this paper greatly.

\noindent{This work is supported by the Melbourne University International Postgraduate Research Scholarship.}

\bsp \label{lastpage}

\end{document}